\documentclass[floatfix,twocolumn,aps,showpacs,showkeys,nofootinbib]{revtex4}

\usepackage{amssymb}
\usepackage{amsmath}
\usepackage{graphics}
\usepackage{epsfig}
\usepackage[dvips]{color}
\usepackage{bm}
\usepackage{color}

\newcommand{\be}{\begin{equation}}
\newcommand{\ee}{\end{equation}}
\newcommand{\bea}{\begin{eqnarray}}
\newcommand{\eea}{\end{eqnarray}}
\newcommand{\beas}{\begin{eqnarray*}}
\newcommand{\eeas}{\end{eqnarray*}}

\newcommand{\nn}{\nonumber\\}

\begin{document}
\title{Phase diagram for charged scalars in a magnetic field at finite temperature}
\author{Alejandro Ayala$^1$, Luis Alberto Hern\'andez$^1$, Jes\'us L\'opez$^1$, Ana J\'ulia Mizher$^1$, Juan Crist\'obal Rojas$^2$, Cristi\'an Villavicencio$^3$}
\affiliation{$^1$Instituto de Ciencias
  Nucleares, Universidad Nacional Aut\'onoma de M\'exico, Apartado
  Postal 70-543, M\'exico Distrito Federal 04510,
  Mexico.\\
  $^2$Departamento de F\'isica, Universidad Cat\'olica del Norte, Casilla 1280, Antofagasta, Chile\\
  $^3$Instituto de Ciencias B\'asicas, Universidad Diego Portales, Casilla 298-V, Santiago, Chile.}

\begin{abstract}

We investigate the nature of the phase transition for charged scalars in the presence of a magnetic background for a theory with spontaneous symmetry breaking. We perform a careful treatment of the negative mass squared as a function of the order parameter and present a suitable method to obtain magnetic and thermal corrections up to ring order for the high temperature limit and the case where the magnetic field strength is larger than the absolute value of the square of the mass parameter. We show that for a given value of the self-coupling, the phase transition is first order for a small magnetic field strength and becomes second order as this last grows. We also show that the critical temperature in the presence of the magnetic field is always below the critical temperature for the case where the field is absent. 

\end{abstract}



\maketitle

\section{Introduction}\label{I}

Magnetic fields appear in several physical systems ranging from the femtoscopic to the astrophysical and cosmological realms and can influence the statistical properties of particles that make up these systems. Peripheral collisions of heavy nuclei at high energies, neutron stars and even the early universe are examples of systems where magnetic fields can help catalyze the deconfinement/chiral restoration~\cite{Ayala1,Agasian:2008tb,Fraga:2008qn,Mizher:2008hf,Mizher:2010zb,Menezes:2008qt, Boomsma:2009yk,Fukushima:2010fe,Johnson:2008vna,Preis:2010cq,Callebaut:2011uc,Avancini:2011zz,Andersen:2011ip,Andersen:2012dz, Skokov:2011ib,Fraga:2012fs,Fukushima:2012xw, Johnson}, the superfluid~\cite{Ayala2} and the electroweak phase transitions~\cite{Ayala3, Simone}, respectively. Lattice simulations have also recently paid attention to the QCD phase structure in the presence of magnetic fields. In this context, it appeared at first that the critical temperature increased with the intensity of the magnetic field \cite{D'Elia:2011zu}. This result agreed with most of the model calculations. Latter results, obtained by considering smaller lattice spacing and physical quark masses, found an opposite behavior \cite{Fodor,Bali:2012zg}. The most recent results, show that such a decrease should be associated to a back reaction of the Polyakov loop, which indirectly feels the magnetic field and drives down the critical temperature for the chiral transition \cite{Bruckmann:2013oba}.

On the other hand, since bosons can condense, they play an important role for the description of phase transitions. When bosons are electrically charged their condensation is also subject to the influence of magnetic fields. However, the field theoretical treatment of charged-boson systems at finite temperature in the presence of a magnetic fields is plagued with subtleties. For example, a na\"\i ve implementation of the condensation condition~\cite{Perez}, whereby the chemical potential is taken to be equal to the ground state energy, leads to a divergence of the particle density of that state. This divergence comes from the effective dimensional reduction of the momentum integrals, since the energy levels separate into transverse and longitudinal (with respect to the magnetic field direction) and the former are given in terms of discrete Landau levels. Thus, the longitudinal mode alone no longer can tame the divergence of the Bose-Einstein distribution unless the system is described in a number of spatial 
dimensions $d$ larger than $4$~\cite{May, Daicic, Elmfors}. This misbehavior can be overcome by a proper treatment of the physics involved when magnetic fields are introduced. For instance, it has recently been shown that even for $d=3$ it is possible to find the appropriate condensation conditions by accounting for the plasma screening effects~\cite{Ayala2}.

Another subtlety --not yet addressed in the literature, to our knowledge-- is found in systems whose vacuum expectation value emerges as a consequence of an spontaneous breaking of symmetry, in the presence of an external magnetic field. When the fields are expanded around the true minimum $v_0$, the squared mass $m^2$ becomes a function of the order parameter $v$ and can become negative for some values in the domain range, $0\leq v\leq v_0$, which is of interest to describe the phase transition at finite temperature.  When not properly treated, these negative values of the squared mass produce a non-analytic behavior of the vacuum energy. Although for the thermal contribution the problem can be avoided by considering a large enough temperature, the purely magnetic field contribution to the vacuum energy, which is important, for instance, for a possible splitting of the chiral and deconfinement transitions as the magnetic field strength increases~\cite{Mizher:2010zb}, requires a proper treatment when $m^2 < 0$. 

In this work we study the nature of the phase transitions for a system of charged scalars influenced by magnetic fields at finite temperature. We show that the system presents first and second order phase transitions as we vary the strength of the self-coupling and of the magnetic field. We address the problem of properly treating the negative mass squared parameter for the vacuum energy as corresponds to a system with spontaneous symmetry breaking. We show that with the appropriate treatment, the vacuum energy is continuous and smooth as a function of $m^2$ when this last transits from positive to negative values. To describe the magnetic field effects we use Schwinger's proper-time method. The work is organized as follows: In Sec.~\ref{II} we compute the vacuum energy to one-loop order for a charged scalar in the presence of a uniform magnetic field as a function of the order parameter. In Sec.~\ref{III} we implement the vacuum stability conditions requiring that to one-loop order, the minimum of the vacuum energy as well as the mass of the scalars do not change from their tree-level values.  We also include thermal effects. We work in the high temperature limit as well as in the limit where the magnetic field strength is larger than the absolute value of the squared mass, so as to avoid having to consider the negative mass squared problem for the matter contribution. We compute the finite temperature effective potential up to the ring diagram contribution and in Sec.~\ref{IV} explore the parameter space looking for the values that produce either a first or second order phase transition. We finally summarize and conclude in Sec.~\ref{concl}.

\section{Vacuum energy}\label{II}

\begin{figure}
\includegraphics[scale=.8]{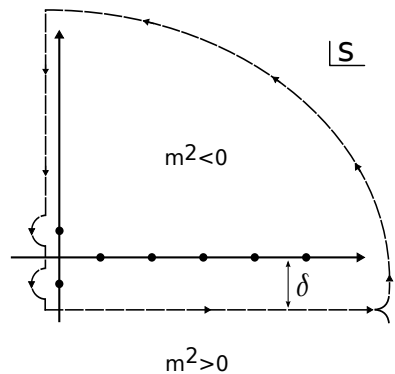}
\caption{Complex $s$-plane. The integration path in Eq.~(\ref{schwprop}) is depicted as a line just below the real axis. Shown also are the poles of the integrand which are the values for $s$ where $\sin (qBs)=0$. These lie on the real axis. The value $s=0$ is a double pole and in order to carry out the integral this pole is splitted. When $m^2>0$ the integration path can be closed in the lower half-plane. However, when $m^2<0$ the integration path needs to be closed in the upper half-plane and therefore it encloses the poles of the integrand.}
\label{fig1}
\end{figure}

When writing a charged scalar field in terms of its real components, the explicit one-loop expression for the vacuum energy density of one these components, in the presence of a uniform magnetic field, is
\bea
   V_B^{(1)} &=& i\int\frac{d^4k}{(2\pi)^4}\ln
   (iD)^{-1/2}\nonumber\\
   &=& - \frac{1}{2}\int dm^2\int\frac{d^4k}{(2\pi)^4}D
\label{VB1}
\eea
where $D$ is the propagator in the presence of a constant magnetic field and $m^2$ is the field's mass squared. For $D$, we use the expression given by Schwinger's proper time method
\bea
   D=\int_{0-i\delta}^{\infty - i\delta} \frac{ds}{\cos (qBs)}
   e^{is\left[k_{\parallel}^2 - k_\perp^2\frac{\tan
   (qBs)}{qBs} - m^2 +i\epsilon\right]},\nonumber\\
\label{schwprop}
\eea
where $k_{\parallel}^2=k_0^2 - k_3^2$ and $k_\perp^2=k_1^2 + k_2^2$ represent the square of the $k^\mu$ components {\it parallel} and {\it transverse}, to the direction of the magnetic field, which is taken in the $\hat{z}$ direction. For convergence, the integration path for the {\it proper time} variable $s$ is taken just below the real axis in the complex $s$-plane. This is implemented by considering that $\delta$ is a small positive quantity. The integration path is depicted in Fig.~\ref{fig1}.

To regulate the ultraviolet divergence, we use dimensional regularization for the longitudinal components which after a Wick rotation are now integrated in $d-2$ dimensions
\bea
   \int\frac{d^4k}{(2\pi)^4}\rightarrow -i
   \Lambda^{4-d}\int\frac{d^{d-2}k_{\parallel}}{(2\pi)^{d-2}}
   \int\frac{d^2k_{\perp}}{(2\pi)^2},
   \label{afterwick}
\eea
where $\Lambda$ is the energy scale for the renormalization, the momentum $k_\parallel$ is now defined in Euclidian space and we take $k_0\to -ik_4$~\cite{LeBellac}. The transverse and longitudinal integrals give
\bea
   \int\frac{d^2k_{\perp}}{(2\pi)^2}e^{-ik_\perp^2\frac{\tan (qBs)}{qB}}=
   -i\frac{qB}{4\pi}\frac{\cos (qBs)}{\sin (qBs)},
   \label{transint}
\eea
and
\bea
   \int\frac{d^{d-2}k_{\parallel}}{(2\pi)^{d-2}}e^{-ik_\parallel^2s}=
   \frac{1}{(4\pi)^{\frac{d-2}{2}}}\frac{1}{(is)^{\frac{d-2}{2}}},
   \label{paralint}
\eea
respectively. The longitudinal integral converges for {\rm Re}$(d)>2$. Also, both integrals converge since $s$ has a small negative imaginary part along its integration path. The vacuum energy density is thus given by
\bea
   V_B^{(1)} &=&  \frac{1}{2}\frac{qB}{4\pi}
\frac{\Lambda^{4-d}}{(4\pi)^{\frac{d-2}{2}}} 
   \int dm^2\nonumber\\
   &\times&\int_{0-i\delta}^{\infty - i\delta} \frac{ds}{\sin (qBs)}
   \frac{e^{-is(m^2 -i\epsilon )}}{(is)^{\frac{d-2}{2}}}.
   \label{vacenergyafterkint}
\eea
The integral over $s$ in Eq.~(\ref{vacenergyafterkint}) can be performed by closing the path and resorting to the residue theorem. When $m^2>0$ the path can be closed on the lower half-plane. However, when $m^2<0$, the path should be closed on the upper half-plane. Therefore, in the latter case the integration contour encloses the poles of $\sin (qBs)^{-1}$ which are located on the real axis. If the last leg of the closing path goes along the imaginary axis, the poles at $s=0$ are located along the path and need to also be properly handled. Let us thus consider in detail the case for $m^2<0$.

\begin{figure}[t!]
\begin{center}
\includegraphics[scale=.9]{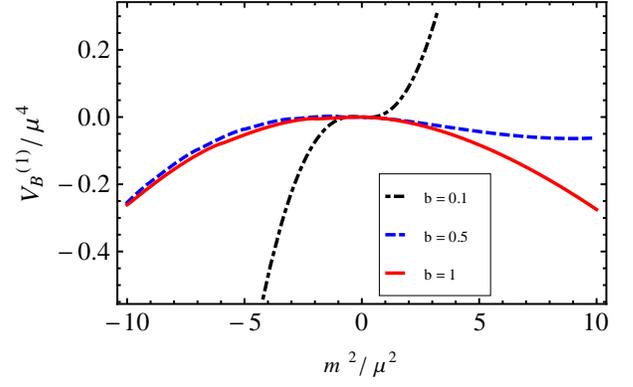}
\end{center}
\caption{Color on-line. One-loop vacuum energy in the presence of the magnetic field expressed in units of the mass parameter $\mu^4$ as a function of the particle mass squared in units of $\mu^2$ for three different values of the magnetic field strength in units of $\mu^2$, $b=qB/\mu^2$. Notice that the curves are continuous and smooth as $m^2$ transits from negative to positive values.}
\label{fig2}
\end{figure}

Using the integration contour that closes in the upper half-plane, depicted in Fig.~\ref{fig1}, we can write
\bea
   &&\oint\frac{ds}{\sin (qBs)}
   \frac{e^{-is(m^2 -i\epsilon )}}{(is)^{\frac{d-2}{2}}}\equiv I_1 + I_2 + I_3\nonumber\\
   &=&2\pi i \sum_{\mbox{\small{res}}}
   \frac{1}{\sin (qBs)}
   \frac{e^{-is(m^2 -i\epsilon )}}{(is)^{\frac{d-2}{2}}},
   \label{countourup}
\eea
where $I_1$ is the integral in Eq.~(\ref{vacenergyafterkint}), namely
\bea
  I_1=\int_{0-i\delta}^{\infty - i\delta} \frac{ds}{\sin (qBs)}
  \frac{e^{-is(m^2 -i\epsilon )}}{(is)^{\frac{d-2}{2}}},
  \label{originalint} 
\eea
$I_2$ is the integral along the quarter circle at infinity, which vanishes, and $I_3$ is given by
\bea
   I_3=\int_{i\infty}^{0-i\delta} \frac{ds}{\sin (qBs)}
  \frac{e^{-is(m^2 -i\epsilon )}}{(is)^{\frac{d-2}{2}}}.
  \label{I3}
\eea
The ultraviolet divergence is contained in $I_3$, so, for the sum over the residues we can take $d=4$. Also, the poles are located at $s=n\pi /qB$. Therefore we can write
\bea
   \sum_{\mbox{\small{res}}}
   \frac{e^{-ism^2}}{(is)\sin (qBs)}
   &=&\frac{1}{i\pi}
   \sum_{n=1}^\infty\frac{(-1)^n}{n}e^{-i\frac{n\pi m^2} 
   {qB}}\nn
   &=&-\frac{1}{i\pi}\ln\left[e^{-i\frac{\pi}{2}(\frac{m^2}{qB}+1)}
   \right.\nn
   &\times&\left.
   \left(e^{i\frac{\pi}{2}(\frac{m^2}{qB}+1)} - 
   e^{-i\frac{\pi}{2}(\frac{m^2}{qB}+1)}\right)\right]\nn
   &=&-\frac{1}{i\pi}\left[\ln (2) + \ln\left|\cos\left(\frac{\pi m^2}{2qB}\right)\right|\right.\nn
   &-&\left.\frac{i\pi m^2}{2qB}\right],
   \label{sumres}
\eea
where $m^2$ is to be understood as $m^2-i\epsilon$. To compute $I_3$ we make the change of variable $s=i\tau$ to write
\bea
   I_3&=&-\frac{1}{(i)^{d-2}}\int_0^\infty\frac{d\tau}{\sinh (qB\tau)}
   \frac{e^{\tau m^2}}{(\tau)^\frac{d-2}{2}}+I_0\nn
   &=&2\sum_{l=0}^\infty
   \int_0^\infty\frac{d\tau}{(\tau)^\frac{d-2}{2}}
   e^{\tau [m^2-(2l+1)qB]}+I_0\nn 
   &=&2\Gamma (2-\frac{d}{2})\sum_{l=0}^\infty
   \frac{1}{[(2l+1)qB-m^2]^{2-\frac{d}{2}}}+I_0\nn
   &=&2
   \frac{\Gamma (2-\frac{d}{2})\zeta (2-\frac{d}{2},\frac{1}{2} - 
   \frac{m^2}{2qB})}{(2qB)^{2-\frac{d}{2}}}+I_0,
   \label{I3expl}
\eea
where $\Gamma$ and $\zeta$ are the gamma and Hurwitz zeta functions, respectively and we have taken $d=4$ in the factor $(i)^{d-2}\rightarrow -1$. $I_0$ is the contribution from the double pole at $s=0$. To compute this last piece, we first take $d=4$, since the ultraviolet divergence is already contained in the first piece of $I_3$, and then split the double pole at $s=0$ such that
\bea
   \frac{1}{(\tau)\sinh (qB\tau)}\rightarrow
   \frac{1}{(\tau-\delta')\sinh [qB(\tau +\delta')]}
   \label{I0}
\eea
where $0<\delta'<\delta$. Since the poles lie along the integration path, we use Cauchy's prescription and thus
\bea
   I_0&=&-\frac{i\pi}{i^2}
   \sum_{\mbox{\small{res}}}
   \left[\frac{1}{\sin (qB\tau)}
   \frac{e^{\tau m^2}}{(\tau)}\right]\nn
   &\rightarrow&
   i\pi \lim_{\delta'\rightarrow 0}
   \left[\frac{e^{m^2\delta'}}{\sinh (2qB\delta')} - \frac{e^{-m^2\delta'}}{2qB\delta'}\right]\nn
   &=&\frac{i\pi m^2}{qB}.
   \label{splitpoles}
\eea
Bringing together Eqs.~(\ref{countourup}),~(\ref{sumres}),~(\ref{I3expl}) and~(\ref{splitpoles}) into Eq.~(\ref{vacenergyafterkint}) and rearranging the factors, we get
\bea
   V_B^{(1)} &=& -\frac{1}{2}\frac{2qB}{(4\pi)^2}
\left(\frac{4\pi\Lambda^2}{2qB}\right)^{\frac{4-d}{2}}   
   \int dm^2\nonumber\\
   &&\left[\ln (2) + \ln\left|\cos\left(\frac{\pi m^2}{2qB}\right)\right|\right.\nn
   &+&\left.\Gamma \left(\frac{4-d}{2}\right)\zeta \left(\frac{4-d}{2},\frac{1}{2} - 
   \frac{m^2}{2qB}\right)\right],
   \label{V1d}
\eea
where in the coefficients of the $\ln$ terms we have already set $d=4$. We now expand around $d=4-2\varepsilon$, with $\varepsilon\rightarrow 0^+$ to get
\bea
   V_B^{(1)} &=&-\frac{1}{2}\frac{2qB}{(4\pi)^2}
   \left[1 + \varepsilon\ln\left(\frac{4\pi\Lambda^2}{2qB}\right)\right]\int dm^2\nn
   &\times&
   \left\{\ln (2) + \ln\left|\cos\left(\frac{\pi m^2}{2qB}\right)\right|
   +\left[\frac{1}{\varepsilon}-\gamma \right]\right.\nn
   &\times&\left.
            \left[\zeta \left(0,\frac{1}{2} - \frac{m^2}{2qB}\right)
    +      \varepsilon\zeta' \left(0,\frac{1}{2} - \frac{m^2}{2qB}\right)
            \right]\right\},\nn
    \label{Vafterexp}
\eea
where $\gamma$ is Euler's gamma and $\zeta'(0,a)=[d\zeta(x,a)/dx)]_{x=0}$. We now use that $\zeta(0,a)=1/2-a$ and $\zeta'(0,a)=\ln(\Gamma(a)/\sqrt{2\pi})$. Dropping the non-leading terms when $\varepsilon\rightarrow 0$ and rearranging factors, we get
\bea
   V_B^{(1)} &=&-\frac{1}{2(4\pi)^2}\int dm^2\nn
   &\times&\left\{m^2\left[\frac{1}{\varepsilon}
   +\ln(4\pi )-\gamma
   +\ln\left(\frac{\Lambda^2}{2qB}\right)\right]\right.\nn
   &+&2qB\left[\ln\left|\cos\left(\frac{\pi m^2}{2qB}\right)\right|\right.\nn
   &+&\left.\left.
   \ln\left(\Gamma\left(\frac{1}{2}-\frac{m^2}{2qB}\right)\sqrt{\frac{2}{\pi}}\ \right)\right]\right\}.
   \label{Vaftermu}
\eea
A convenient choice for the renormalization scale is $\Lambda=\mu e^{-1/2}$, where the mass scale $\mu$ is the mass appearing in the scalar Lagrangian before symmetry breaking. With this choice we have
\bea
   V_B^{(1)} &=&-\frac{1}{2(4\pi)^2}\int dm^2
m^2\left\{\frac{1}{\varepsilon} + \ln (4\pi ) - \gamma\right.\nn
   &-&1 + \ln\left(\frac{\mu^2}{2qB}\right)\nn
   &+&\frac{2qB}{m^2}\left[\ln\left|\cos\left(\frac{\pi m^2}{2qB}\right)\right|\right.\nn
   &+&\left.\left.
   \ln\left(\Gamma\left(\frac{1}{2}-\frac{m^2}{2qB}\right)\sqrt{\frac{2}{\pi}}\ \right)\right]\right\}.   
   \label{Vaftermu}
\eea
The quantity
\bea
   m^2\left[\frac{1}{\varepsilon}+\ln (2\pi ) - \gamma\right],
   \label{renmass}
\eea
can be canceled by the introduction of a suitable counter-term to renormalize the mass squared. Therefore, after mass renormalization we can write
\bea
   \left.V_B^{(1)}\right|_{m^2<0} &=&-\frac{1}{2(4\pi)^2}\int
dm^2 m^2
   \left\{ -1+
   \ln\left(\frac{\mu^2}{qB}\right)\right.\nn
   &+&\frac{2qB}{m^2}\left[\ln\left|\cos\left(\frac{\pi m^2}{2qB}\right)\right|\right.\nn
   &+&\left.\left.
   \ln\left(\Gamma\left(\frac{1}{2}-\frac{m^2}{2qB}\right)\sqrt{\frac{2}{\pi}}\ \right)\right]\right\}.   
   \label{Vaftermassren}
\eea
One can readily check that when $qB\rightarrow 0$, Eq.~(\ref{Vaftermassren}) gives
\bea
   \left.V_B^{(1)}\right|_{m^2<0} \stackrel{qB\rightarrow
0}{\longrightarrow}-\frac{m^4}{(8\pi)^2}
   \left[\frac{1}{2}+\ln\left(\frac{2\mu^2}{-m^2}\right)\right].
   \label{limVmnegB0}
\eea
On the other hand, when $m^2>0$, one can close the integration contour on the lower half-plane, as depicted in Fig.~\ref{fig1}. After mass renormalization the result is
\bea
   \left.V_B^{(1)}\right|_{m^2>0} &=&-\frac{1}{2(4\pi)^2}\int
dm^2 m^2
   \left\{ -1 +
   \ln\left(\frac{\mu^2}{qB}\right)\right.\nn
   &-&\left.\frac{2qB}{m^2}
   \ln\left(\Gamma\left(\frac{1}{2}+\frac{m^2}{2qB}\right)/\sqrt{2\pi}\ \right)\right\}.\nn
   \label{Vaftermassrenmpos}
\eea
When $qB\rightarrow 0$, Eq.~(\ref{Vaftermassrenmpos}) gives
\bea
   \left.V_B^{(1)}\right|_{m^2>0} \stackrel{qB\rightarrow
0}{\longrightarrow}-\frac{m^4}{(8\pi)^2}
   \left[\frac{1}{2}+\ln\left(\frac{2\mu^2}{m^2}\right)\right].
   \label{limVmposB0}
\eea
One can also check the continuity of the expressions when $m^2$ changes from positive to negative values. This is accomplished by taking the limit when $m^2\rightarrow 0$. The result is
\bea  
\left.\frac{V_B^{(1)}}{m^2}\right|_{m^2<0}=\left.\frac{V_B^{
(1)}}{m^2}\right|_{m^2>0}
   \stackrel{m^2\rightarrow
0}{\longrightarrow}-\frac{qB}{(4\pi)^2}\ln (\sqrt{2}).
   \label{limitscoincide}
\eea
Figure~\ref{fig2} shows the vacuum energy, Eqs.~(\ref{Vaftermassren})
and~(\ref{Vaftermassrenmpos}), divided by $\mu^4$, as a function of $x=m^2/\mu^2$, for three different values of the magnetic field strength in units of $\mu^2$, $y=qB/\mu^2$. Notice that the curves are continuous and smooth at $m^2=0$. 

We now proceed to put together the one-loop and the tree-level vacuum energy contributions and to find the thermal corrections. For these purposes, we need to work with an specific Lagrangian and for simplicity we choose to work within the  Abelian Higgs model.

\section{Effective potential}\label{III}

The Abelian Higgs model is given by the Lagrangian~\cite{Das} 
\bea
   {\mathcal{L}}=(D_{\mu}\phi)^{\dag}D^{\mu}\phi+\mu^{2}\phi^{\dag}\phi-\frac{\lambda}{4}   
   (\phi^{\dag}\phi)^{2},
\label{lagrangian}
\eea
where $\phi$ is a charged scalar field and
 \bea
   D_{\mu}\phi=\partial_{\mu}\phi-ieA_{\mu}\phi,
\label{dcovariant}
\eea
is the covariant derivative. $A^\mu$ is the vector potential corresponding to an external magnetic field directed along the $\hat{z}$ axis,
\bea
   A^\mu=\frac{B}{2}(0,-y,x,0).
\label{vecpot}
\eea
The squared mass parameter $\mu^2$ and the self-coupling $\lambda$ are taken to be positive.

We can write the complex field $\phi$ in terms of their real components $\sigma$ and $\chi$,
\bea
   \phi(x)&=&\frac{1}{\sqrt{2}}[\sigma(x)+i \chi(x)],\nn
   \phi^{\dag}(x)&=&\frac{1}{\sqrt{2}}[\sigma(x)-i\chi(x)].
\label{complexfield}
\eea
To allow for an spontaneous breaking of symmetry, we let the $\sigma$ field to develop a vacuum expectation value $v$
\bea
   \sigma \rightarrow \sigma + v,
\label{shift}
\eea
which can later be taken as the order parameter of the theory. After this shift, the Lagrangian can be rewritten as
\bea
   {\mathcal{L}} &=& -\frac{1}{2}[\sigma(\partial_{\mu}-ieA_{\mu})^{2}\sigma]-\frac{1}
   {2}\left(\frac{3\lambda v^{2}}{4}-\mu^{2} \right)\sigma^{2}-\frac{\lambda}{16}\sigma^{4}\nn
   &-&\frac{1}{2}[\chi(\partial_{\mu}-ieA_{\mu})^{2}\chi]-\frac{1}{2}\left(\frac{\lambda v^{2}}{4}-   
   \mu^{2} \right)\chi^{2}-\frac{\lambda}{16}\chi^{4} \nn
  &+&\frac{\mu^{2}}{2}v^{2}-\frac{\lambda}{16}v^{4}+{\mathcal{L}}_{I},
  \label{lagranreal}
\eea
where ${\mathcal{L}}_{I}$ represents the interaction Lagrangian after symmetry breaking. It is well known that for the Abelian Higgs model, with a local, spontaneously broken gauge symmetry, the gauge field $A^\mu$ acquires a finite mass and thus cannot represent the physical situation of a massless photon interacting with the charged scalar field. Therefore, for the discussion we ignore the mass generated for $A^\mu$ and concentrate on the scalar sector. From Eq.~(\ref{lagranreal}) we see that the $\sigma$ and $\chi$ squared masses are given by
\bea
  m^{2}_{\sigma}&=&\frac{3}{4}\lambda v^{2}-\mu^{2},\nn
  m^{2}_{\chi}&=&\frac{1}{4}\lambda v^{2}-\mu^{2}.
  \label{masses}
\eea

\subsection{Tree plus one-loop vacuum energy}

The tree-level potential is given by
\be
  V^{(tree)}=-\frac{1}{2}\mu^{2}v^{2}+\frac{\lambda}{16}v^{4}.
\label{treelevel}
\ee
The minimum is obtained for
\be
  v_{0}=\frac{2\mu}{\sqrt{\lambda}}.
  \label{vefmin}
\ee
Notice that  
\bea
  \frac{d^{2}V^{(tree)}}{dv^{2}}&=&\frac{3\lambda v^{2}}{4}-\mu^{2}\nn
  &=&m_\sigma^2
  \label{condmass}
\eea
and also that the field $\chi$ corresponds to the Goldstone boson.

\begin{figure}[t!]
\begin{center}
\includegraphics[scale=.9]{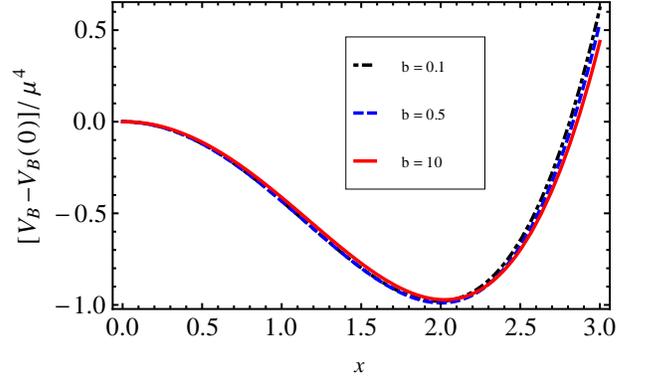}
\end{center}
\caption{Color on-line. Vacuum energy up to one-loop order in the presence of the magnetic field expressed in units of the mass parameter $\mu^4$, as a function of the order parameter in units of $\mu$, $x=v/\mu$, for three different values of the magnetic field strength in units of $\mu^2$, $b=qB/\mu^2$ and for $\lambda=1$. The magnetic field contribution is small even for the largest field intensity $b=10$.}
\label{fig3}
\end{figure}

The vacuum energy up to one-loop level for $qB=0$ can be obtained by combining Eqs.~(\ref{limVmnegB0}),~(\ref{limVmposB0}) and~(\ref{treelevel}) which yields
\bea
   V^{(tree)}+V^{(1)} &=& -\frac{1}{2}\mu^{2}v^{2}+\frac{\lambda}{16}v^{4}\nn
   &-&\frac{m_\sigma^4}{(8\pi)^2}
   \left[\frac{1}{2}+\ln\left|\frac{2\mu^2}{m_\sigma^2}\right|\right]\nn
   &-&\frac{m_\chi^4}{(8\pi)^2}
   \left[\frac{1}{2}+\ln\left|\frac{2\mu^2}{m_\chi^2}\right|\right].
   \label{treeplusoneloop}
\eea
In order that the one-loop correction to the vacuum energy for $qB=0$ preserves the tree-level values of $v_0$ as well as the sigma field mass, we implement the {\it stability conditions} introducing two finite constants $\delta\mu^2$ and $\delta\lambda$ in such a way that
\bea
   V^{(tree)}+V^{(1)} \rightarrow V&=&-\frac{1}{2}\mu^{2}v^{2}+\frac{\lambda}{16}v^{4}\nn
   &-&\frac{m_\sigma^4}{(8\pi)^2}
   \left[\frac{1}{2}+\ln\left|\frac{2\mu^2}{m_\sigma^2}\right|\right]\nn
   &-&\frac{m_\chi^4}{(8\pi)^2}
   \left[\frac{1}{2}+\ln\left|\frac{2\mu^2}{m_\chi^2}\right|\right]\nn
   &-&\frac{\delta\mu^2}{2}v^2 + \frac{\delta\lambda}{16}v^4.
   \label{treeplusonecorrected}
\eea
$\delta\mu^2$ and $\delta\lambda$ are fixed by requiring that
\bea
   \left.\frac{1}{2v}\frac{dV}{dv}\right|_{v=v_0}&=&0\nn
   \left.\frac{d^{2}V}{dv^{2}}\right|_{v=v_0}&=&2\mu^{2},
   \label{conds}
\eea
and the solution is
\bea
   \delta\mu^2&=&-\frac{9\lambda}{(8\pi)^2}\mu^2\nn
   \delta\lambda&=&-\frac{9\lambda^2}{(8\pi)^2}.
\label{condsexpl}
\eea
Including the magnetic field contribution, the vacuum energy, after implementing the stability conditions can be written as
\bea
   V_B&=&-\left(1-\frac{9\lambda}{(8\pi)^2}\right)\frac{\mu^2}{2}v^2
   + \left(1-\frac{9\lambda}{(8\pi)^2}\right)\frac{\lambda}{16}v^4\nn
   &+& \sum_{i=\sigma,\chi}\left[\left.V_B^{(1)}\right|_{m_i^2<0} + 
   \left.V_B^{(1)}\right|_{m_i^2>0}\right],
\label{vaccomplmagfield}
\eea
where $V_B^{(1)}|_{m^2<0}$ and $V_B^{(1)}|_{m^2>0}$ are given by Eqs.~(\ref{Vaftermassren}) and~(\ref{Vaftermassrenmpos}), respectively. Figure~\ref{fig3} shows the vacuum energy, Eq.~(\ref{vaccomplmagfield}), in units of the mass parameter $\mu^4$ as a function of the order parameter, in units of $\mu$, $x=v/\mu$, for three different values of the magnetic field strength, in units of $\mu^2$, $b=qB/\mu^2$, for $\lambda=1$. Notice that the magnetic field contribution to the vacuum energy is rather small, even for the largest field intensity, $b=10$, considered. It's effect is a slight shift of the classical minimum toward larger values as compared to the case where no magnetic field is applied. 

\subsection{One-loop thermal corrections} 
In order to find the finite temperature contribution, we work in the imaginary-time
formulation of thermal field theory. The integration over the momentum components is
carried out in Eucledian space, as in Eq.~(\ref{afterwick}), where the energy
takes on discrete values, namely $-ik_0=\omega_n=2n\pi T$~\cite{LeBellac}, as corresponds to a Matsubara frequency for bosons, with $n$ an integer,
\bea
   \int\frac{d^4k}{(2\pi)^4}\rightarrow
   iT\sum_n\int\frac{d^3k}{(2\pi)^3}.
\label{intsummats}
\eea
Thus, the one-loop contribution to the effective finite-temperature potential in the presence of the magnetic field is given by
\bea
   V^{(1)}_{B,T}&=&\sum_{i=\sigma,\chi}\frac{T}{2}\sum_n\int\frac{d^3k}{(2\pi)^3}
   \ln [\Delta_{B,T} (\omega_n, k;m^2_i)^{-1}],\nn
   &=&
   \sum_{i=\sigma,\chi}\frac{T}{2}
   \sum_n\int\frac{d^3k}{(2\pi)^3}
   \int dm_i^2\Delta_{B,T} (\omega_n, k;m^2_i),\nn
\label{finiteT1}
\eea
where the Matsubara propagator in the presence of the magnetic field is given by
\bea
   \Delta_{B,T}(\omega_n, k;m^2_i)=iD(k_0=i\omega_n,k;m^2_i),
\label{matsprop}
\eea
with $D(k_0,k;m^2)$ as in Eq.~(\ref{schwprop}). We work explicitly in the high temperature limit, namely,
\bea
   T^2\gg |m^2_i|\ , qB,
\label{highTlim}
\eea
but up to this point we do not restrict the strength of $qB$ compared to $|m^2|$. For the $n\neq 0$ modes in Eq.~(\ref{finiteT1}) one can resort to expanding the Matsubara propagator in powers of $qB/ T^2$, in the same fashion as in Ref.~\cite{Ayala4}. Nevertheless, for $n=0$, use of this approximation would amount to restricting ourselves to the situation where $qB \ll |m^2_i|$. To avoid such limitation, we treat the zero frequency separately. In this way
\begin{eqnarray}
   \sum_n\Delta_{B,T}(\omega_n, k;m_i^2) &=& 
   \sum_{n\neq 0}\Delta_{B,T}(\omega_{n\neq 0}, k;m_i^2)\nn 
   &+&
   \Delta_{B,T}(\omega_{n=0}, k;m_i^2),
   \label{modeseparation}
\end{eqnarray}
where for the first term on the right-hand side of Eq.~(\ref{modeseparation}) we use  the weak field expansion as in Ref.~\cite{Ayala4}, namely
\begin{eqnarray}
\Delta_{B,T}(\omega_{n\neq 0}, k;m_i^2) &\approx& 
   \frac{1}{\omega_n^2+k^2+m_i^2}\nn
   &\times&\left[ 1 - \frac{(qB)^2}
   {(\omega_n^2+k^2+m_i^2)^2}\right.\nn
   &+&
   \left.\frac{2(qB)^2k_\perp^2}{(\omega_n^2+k^2+m_i^2)^3}\right],
   \label{nonzeromode}
\end{eqnarray}
and for the second one we keep the Schwinger proper time expression in Eucledian space for $n=0$, that is
\begin{eqnarray}
\Delta_{B,T}(\omega_{n=0}, k;m_i^2)&=&
 \int_{0-i\delta}^{\infty - i\delta} \frac{ds}{\cos (qBs)}\nn
   &\times&e^{-is\left[k_{z}^2 + k_\perp^2\frac{\tan
   (qBs)}{qBs} + m_i^2 -i\epsilon\right]}.\nn
\label{zeromode}
\end{eqnarray}
Inserting Eqs.~(\ref{zeromode}) and~(\ref{nonzeromode}) into Eq.~(\ref{finiteT1}) we obtain
\bea
   V_{B,T}^{(1)}=V_{B,T}^{(1a)} + V_{B,T}^{(1b)}
\label{finiteT2}
\eea
where
\bea
  V_{B,T}^{(1a)}&\equiv&\frac{T}{2}\sum_{i=\sigma, \chi}\sum_{n\neq 0}\int \frac{d^3 k}{(2\pi)^3}
  \int dm_i^2 \Delta_{B,T}(\omega_{n\neq 0},k;m_i),\nn
  \label{nonzero_mode_pot}
\eea
and
\bea
  V_{B,T}^{(1b)}&\equiv&\frac{T}{2}\sum_{i=\sigma, \chi}\int \frac{d^3 k}{(2\pi)^3}
  \int dm_i^2 \Delta_{B,T}(\omega_{n=0},k;m_i).\nn
  \label{zero_mode_pot}
\eea
We can explicitly carry out the sum and integrals in Eq.~(\ref{nonzero_mode_pot}). The sum over the non-zero modes is performed by means of the Mellin technique~\cite{Bedingham}. To compute the integrals in Eq.~(\ref{zero_mode_pot}), care must be taken for the explicit evaluation of $\Delta_{B,T}(\omega_{n=0}, k;m_i^2)$ since, in order to avoid the subtleties associated to negative values of $m^2$, the combination $m_i^2 + (2l+1)qB$ must be positive. This can be achieved by requiring that $qB > |m^2|$. Notice that with this choice, hereby the hierarchy of scales we work with is explicitly
\bea
   T^2\gg qB > |m_i^2|.
\label{hierarchy}
\eea
Under these conditions we get
\begin{eqnarray}
   V_{B,T}^{(1a)}&=&\sum_{i=\sigma, \chi}\left\{
   -\frac{T^4\pi^2}{90}
   +\frac{T^2m_i^2}{24}\right.\nn 
   &-&
   \frac{m_i^4}{64\pi^2}\left[\ln\left(\frac{|m_i^2|}{(4\pi T)^2}\right) +2\gamma-\frac{3}{2}\right]\nn
   &+& \frac{(qB)^2}{48\pi}\left[\ln\left(\frac{\mu}{2\pi T}\right) + \gamma + \frac{1}{2}\right]\nn
   &-& (qB)^2\left[\frac{\zeta (3)}{192\pi^2}\left(\frac{m_i}{2\pi T}\right)^2
   + \frac{\zeta (5)}{256\pi^2}\left(\frac{m_i}{2\pi T}\right)^4\right]\nn 
   &+& \left.{\mathcal{O}}[(m_i^2/T^2)^3]\right\},
\label{omegaT}
\end{eqnarray}
and
\begin{eqnarray}
   V_{B,T}^{(1b)}=\sum_{i=\sigma, \chi}
   \frac{T(2qB)^{3/2}}{8\pi}\zeta\left(-\frac{1}{2},\frac{m_i^2+qB}{2qB}\right),
\label{omegaTB}
\end{eqnarray}
where we have subtracted the vacuum contribution and the mass renormalization, as these contributions are already taken care of in Eqs.~(\ref{Vaftermassren}) and~(\ref{Vaftermassrenmpos}). To write Eq.~(\ref{omegaT}), we have also subtracted another infinite piece associated to the charge renormalization, namely
\bea
(qB)^2\left[\frac{1}{\varepsilon}+\ln (4\pi ) - \gamma\right].
\label{chrgerenor}
\eea
\begin{figure}[t!]
\begin{center}
\includegraphics[scale=0.8]{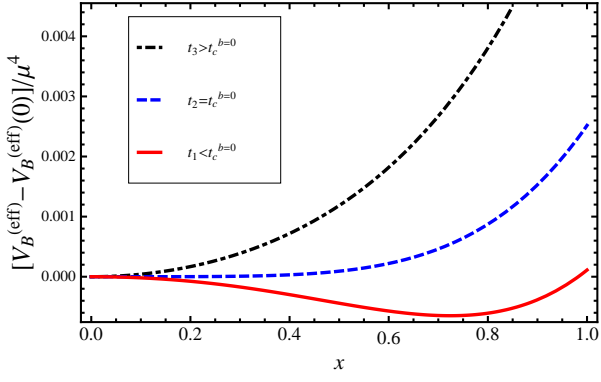}
\end{center}
\caption{Color on-line. Effective potential in units of $\mu^4$ for $qB=0$, $\lambda=0.1$ and three values of $t=T/ \mu$ as a function of $x=v/\mu$.}
\label{fig4}
\end{figure}

\subsection{Ring diagrams}

The ring contribution to the effective potential is given by~\cite{LeBellac}
\bea
   V^{({\mbox{\small{ring}}})}_{B,T}&=&\frac{T}{2}\sum_{i=\sigma, \chi}
   \sum_{n}\nn
   &\times&\int \frac{d^3 k}{(2\pi)^3}\ln[1+\Pi (m_i)
   \Delta_{B,T}(\omega_n, k;m_i^2)],\nn
   \label{ring}
\eea
where for the self-energy $\Pi (m_i)$ we take the dominant contribution in the high temperature limit.
\bea
   \Pi (m_i)\equiv \Pi=\lambda\frac{T^2}{12}.
   \label{selfenergy}
\eea
Notice that upon this choice, the self-energy is mass independent. It is well known that in order to account for the leading plasma screening effects, it is enough to just take the $n=0$ Matsubara frequency for the ring contribution~\cite{LeBellac}. Let us furthermore consider the approximation where the self-energy is small so as to expand the argument of the logarithm inside the integrand in Eq.~(\ref{ring}) to yield
\begin{figure}[t!]
\begin{center}
\includegraphics[scale=0.8]{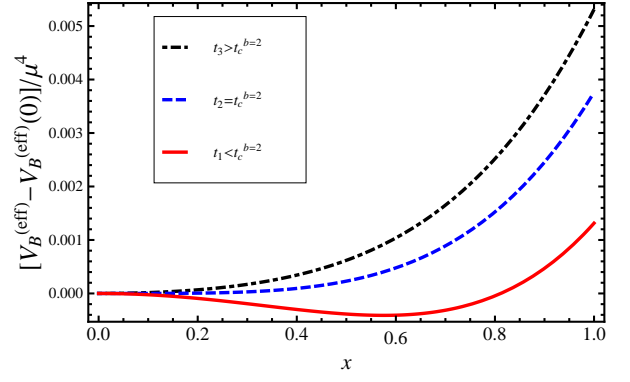}
\end{center}
\caption{Color on-line. Effective potential in units of $\mu^4$ for $b\equiv qB/\mu^2=2$,  $\lambda=0.1$ and three values of $t=T/\mu$ as a function of $x=v/\mu$. For the chosen parameters the phase transitions is second order.}
\label{fig5}
\end{figure}
\begin{figure}[b!]
\begin{center}
\includegraphics[scale=0.8]{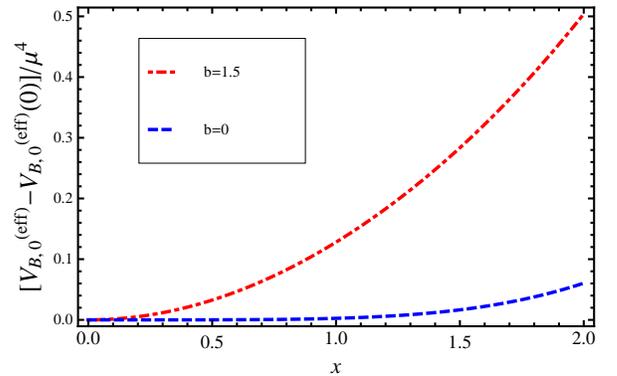}
\end{center}
\caption{Color on-line. Comparison between the effective potentials for $b\equiv qB/\mu^2=0$ and $b=1.5$ for $\lambda=0.1$ evaluated at the critical temperature for the $b=0$ case.}
\label{fig6}
\end{figure}
\bea
   V^{({\mbox{\small{ring}}})}_{B,T}&\simeq&\frac{T}{2}\ \Pi\sum_{i=\sigma, \chi}
   \int \frac{d^3 k}{(2\pi)^3}
   \Delta_{B,T}(\omega_{n=0}, k;m_i^2)\nn
   &=& \Pi\sum_{i=\sigma, \chi}
   \frac{T(2qB)^{1/2}}{8\pi}\zeta\left(\frac{1}{2},\frac{m_i^2 + qB}{2qB}\right).
\label{ringapprox}
\eea

\section{Parameter space}\label{IV}

The complete effective finite temperature potential in the presence of a magnetic field in the high temperature limit, up to the ring contribution, is obtained by adding up Eqs.~(\ref{vaccomplmagfield}),~(\ref{finiteT2}) and~(\ref{ringapprox}), namely
\bea
   V_B^{({\mbox{\small{eff}}})}=V_B + V_{B,T}^{(1)} + V^{({\mbox{\small{ring}}})}_{B,T}.
\label{effpot}
\eea
\begin{figure}[t!]
\begin{center}
\includegraphics[scale=0.83]{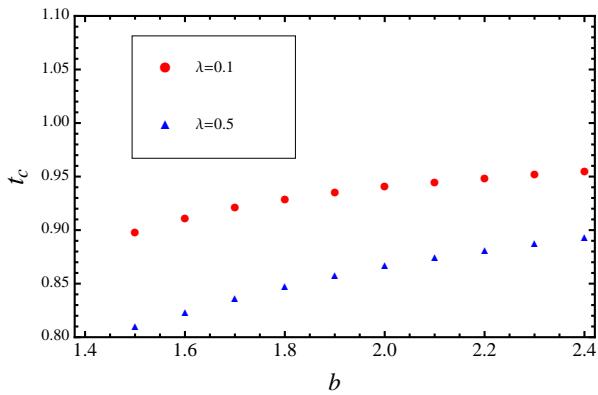}
\end{center}
\caption{Color on-line. Ratio of the $b\equiv qB/\mu^2$ dependent critical temperature divided by the critical temperature for $b=0$ for $\lambda =0.1,\ 0.5$.}
\label{fig7}
\end{figure}
Figure~\ref{fig4} shows the effective potential for $qB=0$, $V^{({\mbox{\small{eff}}})}_{0}$, in units of $\mu^4$ as a function of $x=v/\mu$ for three different temperatures in units of $\mu$ for a fixed value $\lambda=0.1$. Notice that for the conditions that the calculation is valid, namely $qB > |m_i^2|$, we cannot take the limit $qB\rightarrow 0$ straight from Eq.~(\ref{effpot}). Instead, for $qB=0$ we use the well know expression for the effective potential at finite temperature up to the ring diagrams contribution, given by~\cite{Ayala4} 
\bea
   V^{({\mbox{\small{eff}}})}_{0}&=&-\left(1-\frac{9\lambda}{(8\pi)^2}\right)\frac{\mu^2}{2}v^2
   + \left(1-\frac{9\lambda}{(8\pi)^2}\right)\frac{\lambda}{16}v^4\nn
   &+& \sum_{i=\sigma, \chi}\left\{ - \frac{m_i^4}{64\pi^2}\left(\frac{1}{2} 
   + \ln\left[\frac{2\mu^2}{(4\pi T)^2}\right]\right) - \frac{\pi^2T^4}{90} \right.\nn
   &+&\left. \frac{T^2m_i^2}{24} 
   - \frac{T}{12\pi} \left( m_i^2 + \Pi \right)^{3/2} \right\}.
\label{Veff0}
\eea
Figure~\ref{fig5} shows the effective potential for three values of $T$ in units of $\mu$ computed from Eq.~(\ref{effpot}) for $qB/\mu^2=2$ and a fixed value $\lambda=0.1$. Notice that the phase transition for finite $qB$ is delayed with respect to the $qB=0$ case. This is best noticed in Fig.~\ref{fig6} where we compare the effective potential for the $qB=0$ and $qB\neq 0$ cases for the same temperature, which is chosen as the critical temperature for the $qB=0$ case and for $\lambda=0.1$.

\begin{figure}[b!]
\begin{center}
\includegraphics[scale=0.83]{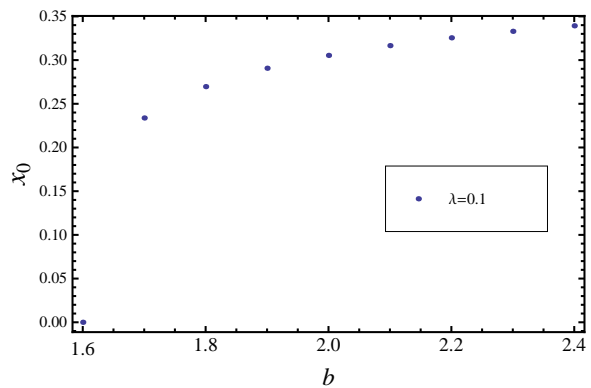}
\end{center}
\caption{Color on-line. The $b\equiv qB/\mu^2$ dependent condensate $v_0$ in units of $\mu$, $x_0=v_0/\mu$ for $\lambda =0.1$.}
\label{fig8}
\end{figure}
For the chosen values of $qB$ and $\lambda$ we notice that the phase transition is second order. As the magnetic field strength increases, the critical temperature grows but remains below the critical temperature for the $qB=0$ case. This is shown in Fig.~\ref{fig7} for two values of $\lambda =0.1,\ 0.5$. Also $v_0$, the value of the order parameter where the effective potential has its minimum in the broken phase, grows with the magnetic field strength. This is shown in Fig.~\ref{fig8} where we plot $x_0\equiv v_0/\mu$ as a function of $b$ for $\lambda =0.1$ and $T$ taken as the critical temperature for the lowest chosen value of $b=1.6$. 
\begin{figure}[t!]
\begin{center}
\includegraphics[scale=0.8]{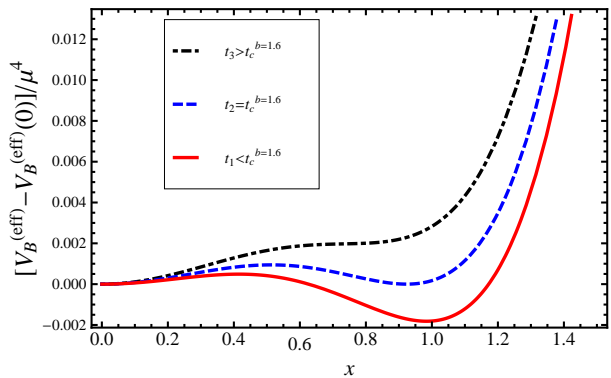}
\end{center}
\caption{Color on-line. Effective potential in units of $\mu^4$ for $b\equiv qB/\mu^2=1.6$,  $\lambda=0.5$ and three values of $t=T/\mu$ as a function of $x=v/\mu$. For the chosen parameters the phase transitions is weakly first order.}
\label{fig9}
\end{figure}
\begin{figure}[b!]
\begin{center}
\includegraphics[scale=0.8]{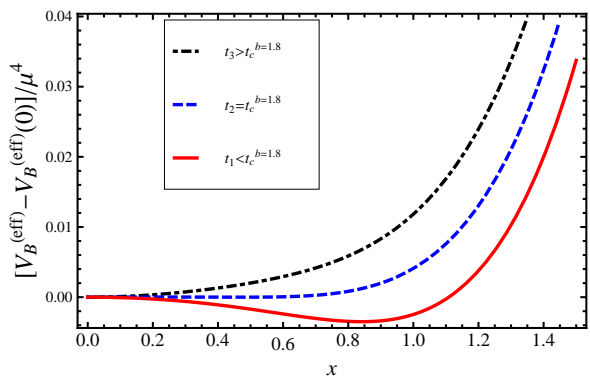}
\end{center}
\caption{Color on-line. Effective potential in units of $\mu^4$ for $b\equiv qB/\mu^2=1.8$,  $\lambda=0.5$ and three values of $t=T/\mu$ as a function of $x=v/\mu$. For the chosen parameters the phase transitions is second order.}
\label{fig10}
\end{figure}
We also find that for certain combinations of $\lambda$ and $qB$, the phase transition becomes weakly first order. This is shown in Fig.~\ref{fig9} where we plot the effective potential for three values of $T$ in units of $\mu$, a fixed value of $qB/\mu^2=1.6$ and $\lambda=0.5$. 
However, starting from the case of a first order phase transition, this becomes again second order as the field strength increases. This is shown in Fig.~\ref{fig10} where we plot the effective potential for three values of $T$ in units of $\mu$, a fixed value of $qB/\mu^2=1.8$ and $\lambda=0.5$. This is in contrast with the findings of Ref.~\cite{Duarte}.
The transit between a first and a second order phase transition as the magnetic field strength increases is illustrated in Fig.~\ref{fig11} where we plot the effective potential for three values of  $qB/\mu^2$ computed at their corresponding critical temperatures.

\begin{figure}[t!]
\begin{center}
\includegraphics[scale=0.8]{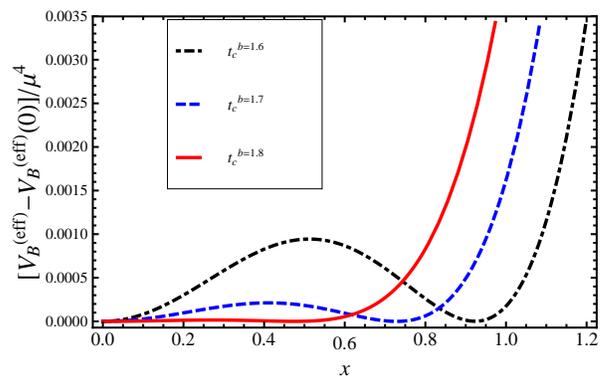}
\end{center}
\caption{Color on-line. Effective potential in units of $\mu^4$ for $\lambda=0.5$ and three values of $b\equiv qB/\mu^2$ at their corresponding critical temperatures as a function of $x=v/\mu$. We observe the transit from a weakly first order to a second order phase transition as the field strength increases.}
\label{fig11}
\end{figure}
\begin{figure}[b!]
\begin{center}
\includegraphics[scale=0.8]{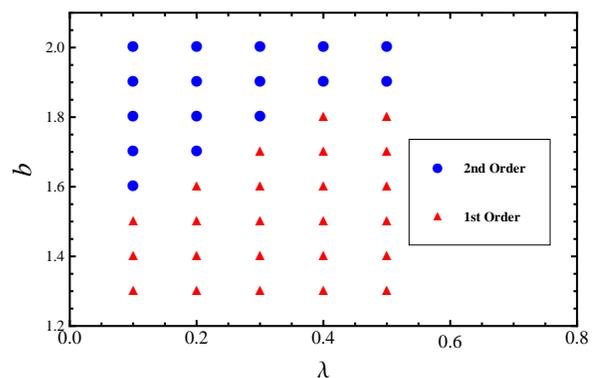}
\end{center}
\caption{Color on-line. Phase diagram for charged scalars in a magnetic field at finite temperature. The horizontal axis shows the self-coupling and the vertical axis the magnetic field sterngth in units of $\mu^2$. The lower right-corner corresponds to first order phase transitions and the upper left-corner to second order phase transitions.}
\label{fig12}
\end{figure}

Figure~\ref{fig12} shows the phase diagram as we vary the self-coupling and the magnetic field strength. Notice that the lower right-corner corresponds to first order phase transitions whereas the upper left-corner corresponds to second order phase transitions.
\\

\section{Summary and conclusions}\label{concl}

In this work we have studied the phase transition at finite temperature for a system made out of charged scalars subject to the effects of a uniform magnetic field. To include the magnetic field in the field theoretical description of the system, we employed Schwinger's proper time method. For the analysis we have computed the finite temperature effective potential up to the ring diagram contribution.  We have worked explicitly within the Abelian Higgs model with spontaneous symmetry breaking and considered the hierarchy of scales such that $T^2 \gg qB > |m^2|$. For the magnetic contribution to the vacuum energy we have made a careful treatment for the case where the square of the mass parameter, as a function of the order parameter, becomes negative. For the matter contribution and for the chosen hierarchy of scales, the subtleties associated with negative values of the square of the mass parameter can be avoided. In this case, we have shown that the system suffers either a first or second order phase 
transition depending on the value of the self-coupling constant $\lambda$ and the strength of the magnetic field $qB$; for a given value of $\lambda$ and a low enough value of $qB > |m^2|$ the phase transition is weekly first order and becomes second order as $qB$ increases. The phase transition gets delayed (the critical temperature is lower), as compared to the case in the absence of a magnetic field, and this increases as the field strength grows. The value for the order parameter that describes the condensate also increases with increasing magnetic field strength. 

In conclusion, we have shown that the phase diagram for a charged scalar system in the presence of a magnetic field has a richer than anticipated structure. However, we emphasize that in order to complete the parameter space studies, a proper handle of the case where $qB<|m^2|$ is required, as well as an extension of the method to lower temperatures. This is work that we are currently pursuing and will be reported elsewhere. 

\section*{Acknowledgments}

Support for this work has been received in part from DGAPA-UNAM under grant number PAPIIT-IN103811, CONACyT-M\'exico under grant number 128534 and FONDECYT under grant  numbers 1130056 and 1120770.

\end{document}